# Phenomenological model for spectral broadening of incoherent light in fibers via self-phase modulation and dispersion

Qinghua Li,[1,2,3] Haitao Zhang,[1,*] Xinglai Shen,[1] He Hao,[1] and Mali Gong[1]

[1] Center for Photonics and Electronics, State Key Laboratory of Precision Measurement and Instruments, Department of Precision Instruments, Tsinghua University, Beijing, 100084, PR China
[2] Department of Engineering Physics, Tsinghua University, Beijing 100084, People's Republic of China
[3] Institute of Applied Electronic, China Academy of Engineering Physics, Mianyang, Sichuan 621900, People's Republic of China

[*]Corresponding author: zhanghaitao@mail.tsinghua.edu.cn

**Abstract:**

A phenomenological model for spectral broadening of incoherent light in silica fibers via self-phase modulation and dispersion is presented, aiming at providing a qualitative and readily accessible description of incoherent light spectral broadening. In this model, the incoherent light is approximated by a cosine power-modulated light with modulation parameters depending on the coherent time and the dispersion in fibers. A simple and practical method for spectral broadening predictions is given, and demonstrated by both the straightforward NLSE-based numerical modeling and series of experiments including narrowband and broadband incoherent light in passive fibers and fiber amplifiers.

Keywords: Kerr effect, self-phase modulation, phenomenological SPM model, partially coherent light, fiber optics, group velocity dispersion

**1. Introduction**

Incoherent lights sources, e.g. the superfluorescence source (SFS) and the superluminescent diode (SLD) source with no longitudinal modes but smoothly distributed photons within the spectral band is preferable for the applications of low coherence interferometry, optical coherence tomography (OCT), and optical gyroscopes [1–3]. Furthermore, the incoherent light may be a promising candidate for the inertial fusion driver, since it is helpful to reduce the Rayleigh Taylor (RT) hydrodynamic instabilities due to its smooth illumination in both time and space domain [4]. If the laser illumination of the imaging lidar is replaced by the incoherent light source, the imaging resolution could be further improved due to the speckle-noise-free feature of the incoherent light [5]. The incoherent light is also an alternative to short-pulse excitation to provide short time resolution, and can even generate coherent ultrashort pulses directly [6].

High power incoherent light can be obtained from fiber amplifiers [7–9]. However, the power scaling of fiber lasers and amplifiers is limited by nonlinear effects, such as stimulated Raman scattering (SRS), stimulated Brillouin scattering (SBS), self-phase modulation (SPM) and so on. Theories and experiments for SRS and SBS of incoherent light have been explored [10, 11]. The SPM and dispersion induced spectrum broadening of incoherent light in fibers can be studied by straightforward solving nonlinear Schrodinger equation (NLSE) [12]. However, the numerically solved output power spectrum is from the integration of NLSE over an ensemble of hundreds of realizations [13], which makes this method somewhat time consuming. In 1990, Manassah put forward an SPM-induced spectral broadening formula for incoherent light without considering dispersion [14, 15]. This formula was verified in the MOPA fiber amplifier seeded by a Q-switched laser, where the dispersion effects were negligible [16]. Under conditions that the dispersion cannot be ignored, e.g. broadband light propagates in fibers for long distance, the dispersion plays a crucial role and eventually balances the nonlinearity [12, 17], and the calculation under the non-dispersion assumption is not proper to give quantitative or even qualitative spectral broadening predictions any more. Considering the dispersion as well as Kerr nonlinearity effects, wave turbulence (WT) [18] was developed to describe the spectral broadening of incoherent light in fibers. WT theory results in the wave turbulence kinetic equation, which is an integro-differential equation deduced from Schrodinger equation under assumption of Gaussian statistics of the fields [19], and is expressed in the form of four-wave mixing (FWM) [13, 20, 21]. In the recent 25 years, WT theory has been successfully applied to conservative nonlinear optical systems [13, 21–23]. However, the predictions of the WT kinetic equation are mainly applicable in the weakly nonlinear regime, i.e., the highly incoherent regime where the strong randomness of the optical wave (small time correlation) makes linear dispersive effects dominant with respect to nonlinear effects [19]. Applied in both strong and weakly nonlinear regimes, to the best of our knowledge, no analytical and mathematical method has been developed.

In this paper, we present a phenomenological model for spectral broadening of incoherent light in silica fibers. This model is applicable in both the strong nonlinear regime and the weakly nonlinear regime because there is no need for the assumption of the dominance of Kerr nonlinear effects or dispersions in this model. This model approximates the incoherent light to a cosine power-modulated light with modulation period depending on the coherent time and modulation depth based on the dispersion in fibers. A simple but practical spectral broadening formula is achieved. Both the straightforward NLSE-based numerical modeling and the verification experiments are performed to verify this model. The experiments include narrowband incoherent light in a passive fiber, broadband incoherent light in a passive fiber and narrowband incoherent light in the Yb-doped fiber amplifier. Our model predictions achieve qualitative agreements with the straightforward NLSE-based numerical modeling and the experimental results.

## 2. Phenomenological model for spectral broadening of incoherent light in silica fibers via SPM and dispersion

Unlike the coherent light, the phase of incoherent light is randomly distributed in frequency domain. By setting a power spectrum with phases randomly distributed in $[0, 2\pi]$, and applying the inverse Fourier transform to its complex amplitude in the frequency domain, the temporal complex amplitude and thus the temporal power envelope of the light can be obtained [24].

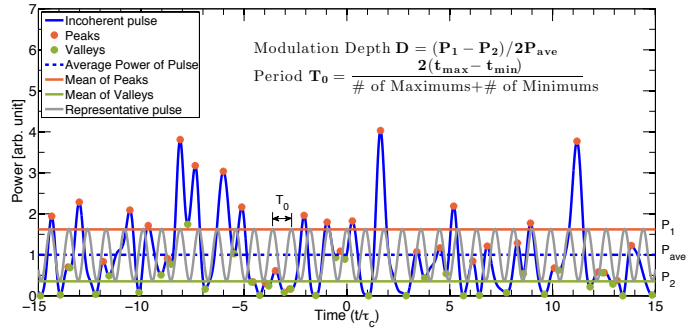

Fig. 1 Typical power envelope (blue curve) and the corresponding simplified envelope (gray line) for incoherent light with Gaussian spectral profile

As shown in Fig. 1, the solid blue curve is the generated incoherent light power envelope according to the above-stated process. The power is normalized by the light average power

$P_{ave}$, and the time is normalized by the light coherent time $\tau_c$, which is calculated by $\tau_c = \lambda^2 / c\Delta\lambda$ for light with central wavelength of $\lambda$ and 3-dB bandwidth of $\Delta\lambda$, and $c$ is the light speed in vacuum. The incoherent light with randomly fluctuated subpulses can be phenomenologically simplified to a regularly modulated pulse with the average power unchanged, as the gray curve shows in Fig. 1, which is a cosine power-modulated pulse with modulation period $T_0$ and modulation depth $D$. The modulation depth $D$ is defined as Eq. (1), where $P_1$ and $P_2$ are the averaged powers of the subpulse peaks and valleys of the original light, and $P_{ave}$ is the average power of light.

$$D = \frac{P_1 - P_2}{2P_{ave}} \tag{1}$$

Subpulse period $T_0$ is defined as the lasting time of the generated incoherent light divided by the number of peak-valley pairs, which is written as Eq. (2).

$$T_0 = \frac{2(t_{max} - t_{min})}{\text{Number of Peaks + Number of Valleys}} \tag{2}$$

Then the original irregular power envelope $P(t)$ can be approximated by $P_{cos}(t)$ as shown in Eq. (3). The modulation period $T_0$ contains the incoherent information of the light, i.e. coherent time $\tau_c$, and the modulation depth $D$ presents the averaged power-fluctuation amplitude of the subpulses. Modulation period and modulation depth both will affect the self-phase modulation strength, and hence the spectral broadening degree.

$$P_{cos}(t) = P_{ave}(1 + D\cos(\frac{2\pi}{T_0}t)) \tag{3}$$

After generating more than 10000 incoherent pulses (like the blue curve in Fig. 1) and doing calculations of Eq. (1) and Eq. (2) for each generation, in a statistical sense, we find $D \approx 0.6235$ and $T_0 \approx 0.955\,\tau_c$ for incoherent light with Gaussian spectrum. For the rectangular-shaped spectrum, $D \approx 0.709$ and $T_0 \approx 1.546\,\tau_c$. It implies the spectrum shapes have considerable impacts on the modulation period and modulation depth. In this paper, we mainly focus on these two shapes, since they are the most common spectral shapes in fiber amplifier systems.

For the fiber with length $L$, the light will experience $L/L_D$ times dispersion-induced total pulse shape changes, where the dispersion length $L_D$ is defined as $L_D = (K_{TB}T_0)^2 / |\beta_2|$. Here, $K_{TB}$ is the transform-limited time-bandwidth product, e.g. $K_{TB} = 0.441$ for Gaussian spectrum and $K_{TB} = 0.443$ for rectangular spectrum. $\beta_2$ is the group velocity dispersion parameter, which is set to $23 \times 10^{-27}\,\text{s}^2/\text{m}$ in calculations for 1 μm wavelength light in silica fibers. Then the light will exhibit $n = L/L_D + 1$ times totally different pulse shapes during its propagation through the fiber. Here the number 1 accounts for the initial pulse shape. We call $n$ the dispersion factor hereinafter.

We will show that the SPM-induced spectral broadening can be greatly weakened if the dispersion factor $n$ is large. If we assume a small part of the light wave of $P(t)$ is currently located at the rising edge of a subpulse, then this part of light will experience SPM-induced downward frequency shift. After one dispersion-induced total pulse shape change, this part of light is possibly located at the falling edge of a subpulse. Then this part of light will experience SPM-induced upward frequency shift, which will cancel out the previous downward frequency shift. The more times the dispersion-induced total pulse shape changes there are, the stronger the cancel-out effect of the frequency shift is. Thus the overall SPM-induced spectral broadening will be greatly weakened with large dispersion factor $n$.

Based on the physical picture described above, by taking the average of $n$ pulses with random phases, an averaged pulse with dispersion-effect-included modulation period $T_0(n)$ and modulation depth $D(n)$ will be obtained. After more than 10000 times calculations, we find in a statistical sense that the modulation period $T_0(n)$ is shown to be independent with $n$ as we expected, and relate with $\tau_c$ as Eq. (4). As indicated by the words in the brackets, the first and second lines in the equations are for incoherent light with Gaussian and rectangular spectra, respectively.

$$T_0(n) \approx \begin{cases} 0.955\tau_c & \text{(Gaussian)} \\ 1.546\tau_c & \text{(rectangular)} \end{cases} \tag{4}$$

For modulation depth $D(n)$, it is found to relate with the dispersion factor $n$ as Eq. (5), where the dispersion factor $n$ is calculated by Eq. (6). Fig. 2 shows the dependence of $D(n)$ on $n$ for the Gaussian spectrum, where the red hollow squares are for the simulation results and the green curve is the fitting by Eq. (5)(Gaussian) with confidence level of $R^2>0.998$. As shown in Fig. 2, $D(n)$ decreases dramatically for $n<100$, and then decreases slower for larger $n$.

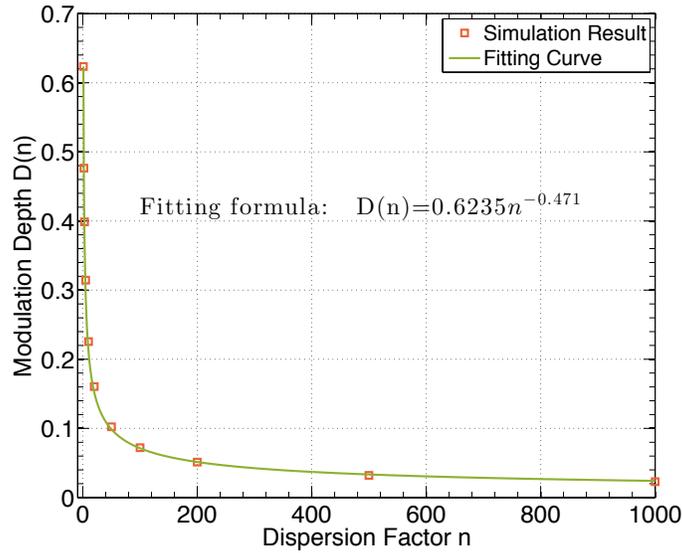

Fig. 2 Dependence of modulation depth $D(n)$ on dispersion factor $n$ for the Gaussian spectrum

$$D(n) \approx \begin{cases} 0.6235n^{-0.471} & \text{(Gaussian)} \\ 0.709n^{-0.471} & \text{(rectangular)} \end{cases} \tag{5}$$

$$n = L/L_D + 1 \approx \begin{cases} 5.638c^2\Delta\lambda^2 L|\beta_2|/\lambda^4 + 1 & \text{(Gaussian)} \\ 2.151c^2\Delta\lambda^2 L|\beta_2|/\lambda^4 + 1 & \text{(rectangular)} \end{cases} \quad (6)$$

With dispersion effects included, the simplified pulse is modified as Eq. (7).

$$P_{\cos}(t,n) = P_{ave}(1 + D(n)\cos(\frac{2\pi}{T_0}t)) \quad (7)$$

According to the self-phase modulation term in the nonlinear Schrodinger equation [25], the time-dependent angular frequency difference $\delta\omega(t)$ can be calculated as Eq. (8), where the effective fiber length is $L_{eff} = (1-\exp(-\alpha L))/\alpha$ for passive fibers with loss coefficient $\alpha$, or $L_{eff} = \int_0^L \exp(g(z)z)dz$ for active fibers with $g(z)$ the gain coefficient at position $z$ of the fiber. The nonlinear coefficient $\gamma$ can be calculated as $\gamma = 2\pi n_2/\lambda A_{eff}$ where nonlinear refractive index $n_2 = 2.6\times 10^{-20}$ m²/W for linearly polarized light with 1 μm wavelength in silica fibers ($n_2 = \frac{5}{6}\times 2.6\times 10^{-20}$ m²/W for random polarization) [26] and $A_{eff}$ is the effective modal area of the fiber. $P(t)$ is the instantaneous pulse power.

$$\delta\omega(t) = -\gamma L_{eff}\frac{\partial}{\partial t}P(t) \quad (8)$$

By setting $P(t) = P_{\cos}(t,n)$ in Eq. (8), we get Eq. (9).

$$\delta\omega(t,n) = \frac{2\pi}{T_0}D(n)\gamma P_{ave}L_{eff}\sin(\frac{2\pi}{T_0}t) \quad (9)$$

Then the 3-dB bandwidth increment $\Delta\lambda_{inc}$ can be calculated by Eq. (10), where $\Delta\omega(n) = 2\overline{|\delta\omega(t,n)|}$ is the angular frequency increment, and $\overline{|\delta\omega(t,n)|}$ means taking time average of the absolute value of $\delta\omega(t,n)$. As we expected, the bandwidth increment is proportional to the modulation depth $D(n)$ and inversely proportional to the modulation period $T_0$.

$$\Delta\lambda_{inc} = \frac{\Delta\omega(n)}{2\pi}\frac{\lambda^2}{c} = \frac{2\overline{|\delta\omega(t,n)|}}{2\pi}\frac{\lambda^2}{c} = \frac{4\lambda^2 L_{eff}\gamma P_{ave}}{\pi c}\frac{D(n)}{T_0} \quad (10)$$

Then spectral broadening ratio can be calculated by Eq. (11), where Eq. (4) and $\tau_c = \lambda^2/c\Delta\lambda$ is used. And the output bandwidth can be calculated by Eq. (12).

$$\frac{\Delta\lambda_{inc}}{\Delta\lambda} \approx \begin{cases} \frac{4}{3}D(n)L_{eff}\gamma P_{ave} & \text{(Gaussian)} \\ \frac{5}{6}D(n)L_{eff}\gamma P_{ave} & \text{(rectangular)} \end{cases} \quad (11)$$

$$\Delta\lambda_{out} = \Delta\lambda(1+\frac{\Delta\lambda_{inc}}{\Delta\lambda}) \approx \begin{cases} \Delta\lambda(1+\frac{4}{3}D(n)L_{eff}\gamma P_{ave}) & \text{(Gaussian)} \\ \Delta\lambda(1+\frac{5}{6}D(n)L_{eff}\gamma P_{ave}) & \text{(rectangular)} \end{cases} \quad (12)$$

Since $D(n)$ in Eq. (11) and Eq. (12) is dependent on the dispersion factor $n$ which is based on the input bandwidth $\Delta\lambda$, Eq. (11) and Eq. (12) are accurate only when the spectral broadening ratio is very small. If the broadening ratio is considerable, e.g. broadening ratio $r = \Delta\lambda_{inc}/\Delta\lambda > 1$, then the dispersion factor $n$ and thus the modulation depth $D(n)$ should be recalculated. By solving NLSE directly, we observe the spectrum is broadened and the modulation period of the light is shortened during its propagation along the fiber. Then, the new modulation period can be modified as $T_0/\Gamma_0$, where $\Gamma_0$ is the average period-shortening factor, which is proportional to the broadening ratio $r$, as shown in Eq. (13). $\alpha$ is the fiber loss for passive fiber, which can be easily replaced by $\alpha = -g$ for gain fiber with gain coefficient $g$. When the fiber loss or the gain is ignored in Eq. (13), i.e. $\alpha = 0$, the period-shortening factor is reduced to $\Gamma_0 = 1 + r/2$.

$$\Gamma_0 = 1 + r\frac{1-(1-e^{-\alpha L})/\alpha L}{1-e^{-\alpha L}} \quad (13)$$

Accordingly, the dispersion length should be replaced by $L_D/\Gamma_0^2$, since the dispersion length is proportional to the second-order of modulation period. Then, the new dispersion factor $n$ can be calculated by Eq. (14).

$$n \approx \begin{cases} 5.638\Gamma_0^2 c^2 \Delta\lambda^2 L|\beta_2|/\lambda^4 + 1 & \text{(Gaussian)} \\ 2.151\Gamma_0^2 c^2 \Delta\lambda^2 L|\beta_2|/\lambda^4 + 1 & \text{(rectangular)} \end{cases} \quad (14)$$

The new modulation depth can be obtained by substituting the new dispersion factor $n$ into Eq. (5). Then Eq. (11) or Eq. (12) can be recalculated to obtain the revised spectral broadening. Although this twice-calculation method is proposed for cases that the spectral broadening ratios are large, it is also suitable for small broadening ratio cases. For the small broadening ratio condition, the twice-calculation method will give predictions quite close to that by the once-calculation method.

It is worth noting that in the above deductions, no assumption of the dominance of Kerr nonlinearities or dispersions is used. Therefore, the spectral broadening formulas deduced above are applicable no matter in the strong nonlinear regime or in the weakly nonlinear regime in principle.

In the following simulations, the twice-calculation method is adopted, and compared with the predictions by directly solving NLSE using split-step Fourier method (SSFM) [25]. The

NLSE-based numerical results are obtained after performing an ensemble average over 200 realizations of the power spectrum of the output field. In the simulations, the light is assumed to be randomly polarized, and has a Gaussian-shaped spectrum. The fiber is the 9/125 passive fiber with effective modal area of $A_{eff}$ =39.5 μm$^2$ for the 1 μm wavelength light, and has a fiber loss of $\alpha$ =0.2 km$^{-1}$, according to the subsequent experiments.

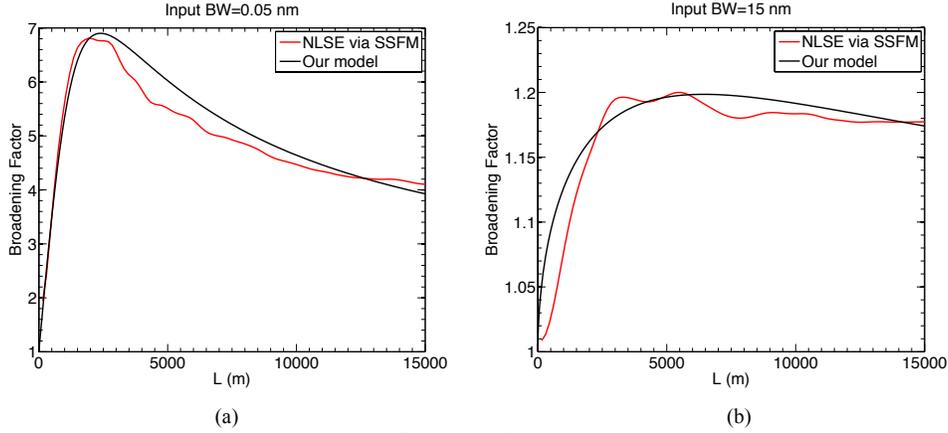

(a) (b)
Fig. 3 Output-to-input bandwidth ratio versus the fiber length for light with input power of 2 W in the 15000 m 9/125 fiber (a) Input BW=0.05 nm (b) Input BW=15 nm

The output-to-input bandwidth ratio versus the fiber length for light with input power of 2 W in the 15000 m 9/125 passive fiber is shown in Fig. 3, where (a) is for input bandwidth of 0.05 nm and (b) is for 15 nm. The red line and back lines are the results by the straightforward solving NLSE via SSFM and by the twice-calculation method, respectively. It shows that our model agrees quantitatively with the straightforward NLSE method for the low power long fiber cases.

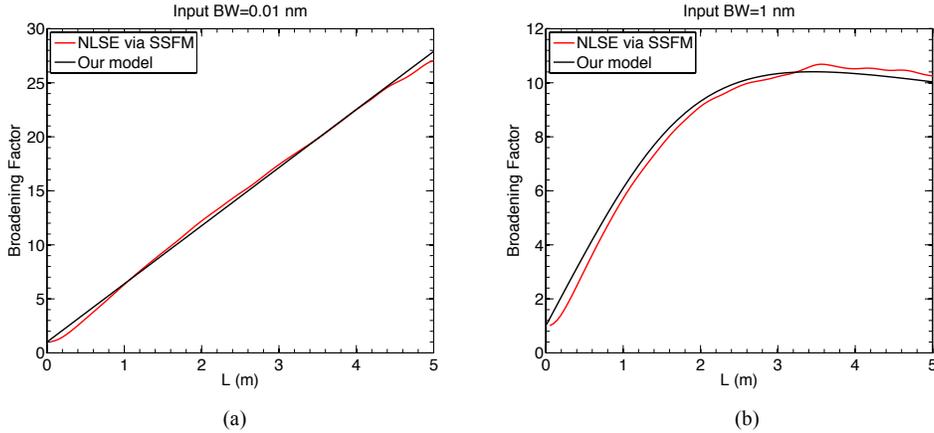

(a) (b)
Fig. 4 Output-to-input bandwidth ratio versus the fiber length for light with input power of 2 kW in the 5 m 9/125 fiber (a) Input BW=0.01 nm (b) Input BW=1 nm

The output-to-input bandwidth ratio versus the fiber length for light with input power of 2 kW in the 5 m 9/125 passive fiber is shown in Fig. 4, where (a) is for input bandwidth of 0.01 nm and (b) is for 1 nm. For the high power short fiber cases, our model also agrees well with the straightforward NLSE method.

Besides, we can see from the above simulations that the twice-calculation method can predict the non-monotonic increase of spectral bandwidth with fiber length. This method will be further verified by experiments in the next section.

## 3. Experiments for spectral broadening of incoherent light in silica fibers

### 3.1 Experiments for passive fibers

Firstly, spectral broadening of the narrowband incoherent light in passive silica fibers is investigated. The experimental setup is shown in Fig. 5. The four-stage single-mode all-fiber amplifier system is seeded by a SLD, with central wavelength of 1051 nm and 3-dB bandwidth of 52 nm. The peak power of the seed is 20 mW. The pulse width is set to 970 ns. Three 8-nm band-pass filters with central wavelength of 1064 nm are placed separately after the SLD, the first- and third-stage amplifiers to narrow down the output spectrum. An AOM is placed after the first stage amplifier to filter out the DC power from the first stage ASE. Two 3-port circulators combined with narrowband high reflective (HR) fiber gratings are placed after the second- and fourth-stage amplifiers, respectively, to further narrow down the pulse spectrum. The first and second fiber gratings are HR with nominal 3-dB bandwidth of 0.126 nm and 0.06 nm centered at 1064.25 nm. The drivers of the SLD, the AOM and all pump LDs are all controlled by a single synchronous controller, so as to realize high efficiency pulsed pumping and amplifications. All the components in the amplifier system are polarization-independent. Therefore, the light is randomly polarized through the whole system. After four-stage amplification, the narrowband signal is sent into a 500 m 9/125 passive silica fiber with a core diameter of 8.2 µm, numerical aperture (NA) of 0.14 and the effective modal area of 39.5 µm$^2$. By comparing the input and output spectral bandwidths of the 500 m 9/125 fiber, spectral broadening for the narrowband incoherent light in passive silica fibers is studied.

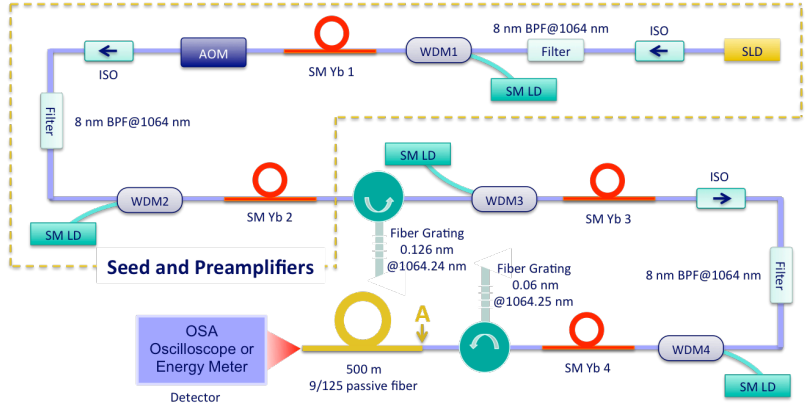

Fig. 5 Experimental setup for spectral broadening of narrowband incoherent light in passive fibers

The fiber loss in the 9/125 passive fiber for 1064 nm light is measured to be $\alpha = 0.2$ km$^{-1}$. The input pulse shapes at different pulse energies are shown as Fig. 6. $E_A$ is the input pulse energy measured at position A as the yellow arrow indicates in Fig. 5. Since the pulse shape remains basically unchanged after propagation through the fiber, the output pulse shapes are not displayed here. With increased pulse energy, the flat-top pulse is getting distorted as the front edge of the pulse is more intensely amplified due to the gain saturation effect. It means the SPM experienced by the front and back part of the pulse are different. Thus the spectral broadenings of each part are different. However, the requirement of nanosecond time resolution is usually too high for the optical spectrum analyzer (OSA) to take instantaneous spectrum. Instead, OSA takes the energy-weighted time-averaged power spectrum $S(\lambda)$ of the pulse $P(t)$ as shown in Eq.(15), where $s(\lambda,t)$ is the instantaneous spectrum of $P(t)$.

$$S(\lambda) = \frac{\int P(t)s(\lambda,t)dt}{\int P(t)dt} \quad (15)$$

Therefore, Eq. (16) is used to calculate the average power of the input pulse $P_{ave}$ for Eq. (12). The integrands in Eq. (16) are integrated in the time section that contains the whole pulse, i.e. $\int P(t)dt$ is the pulse energy. Therefore, Eq. (16) gives the energy-weighted time-average pulse power, which is in accordance with the working principle of the OSA.

$$P_{ave} = \frac{\int P^2(t)dt}{\int P(t)dt} \quad (16)$$

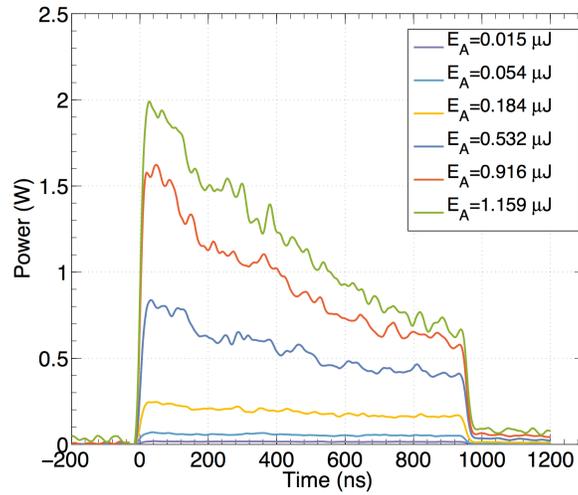

Fig. 6 Input pulse shapes for the 500 m 9/125 fiber at different pulse energies

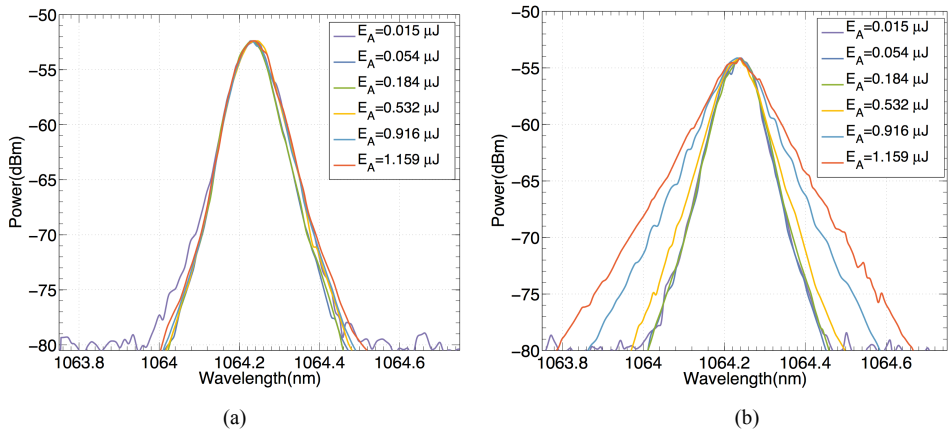

(a)            (b)

Fig. 7 Input (a) and output (b) spectra for the 500 m 9/125 fiber at different pulse energies

The input and output spectra for the 500 m 9/125 fiber at different pump powers of the

4th-stage amplifier are shown in Fig. 7 (a) and (b), respectively. The 3-dB bandwidth of the input pulse is about 0.07 nm, and the output spectrum after 500 m 9/125 passive fiber is broadened. With higher pump power, the output spectrum is broader. In this experiment, $L_{NL} < 1000$ m and $L_D \approx 20000$ m, thus the optical wave is in the strong nonlinear regime ($L_{NL} \ll L_D$). Equation (12)(Gaussian) and related equations are used to calculate the spectral broadenings at different input pulse energies. The experimental input and output bandwidth vs. calculation results by our model are shown in Fig. 8. As shown in Fig. 8, our model agrees qualitatively with the experimental results, with mean errors of less than 16%.

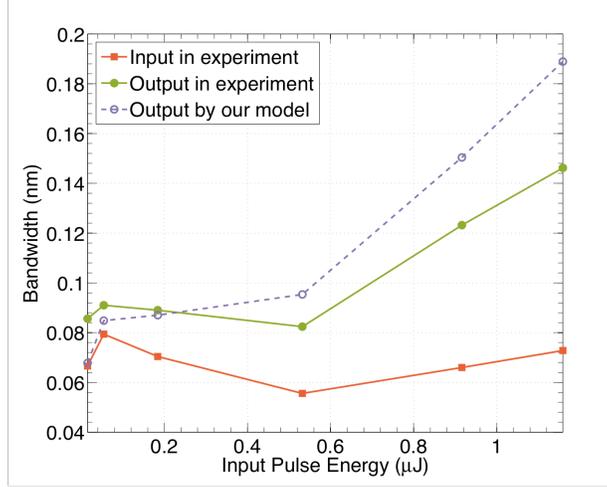

Fig. 8 Experimental results vs. calculation results by our model for the narrowband incoherent light in the 500 m 9/125 passive fiber

The experiments for the broadband incoherent light in passive fibers are also performed. The experimental setup is shown in Fig. 9, where the seed and preamplifiers are the same as those in the first experiment, which include a SLD incoherent light source followed by two SM fiber preamplifiers as the yellow dashed area shows in Fig. 5. The output pulses after pre-amplifications are spectrally shaped by a rectangular-shaped 8-nm band-pass filter with central wavelength of 1064 nm, and then sent into the 1500 m 9/125 passive silica fiber. The relatively longer passive fiber is employed in this broadband experiment so as to achieve obviously spectral broadening.

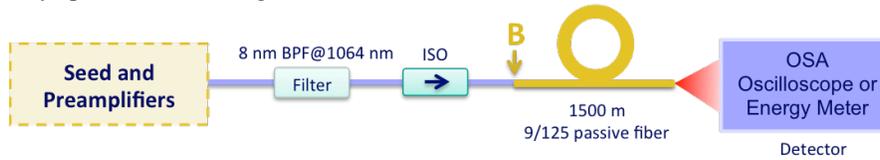

Fig. 9 Experimental setup for spectral broadening of broadband incoherent light in passive fibers

The input pulse shapes for the 1500 m 9/125 fiber at different pulse energies are shown in Fig. 10. $E_B$ is the input pulse energy measured at position B as the yellow arrow indicates in Fig. 9. The light loss is about 25% after the 1500 m propagation. The output pulse shapes have similar features and behaviors, hence not displayed here. Since the gain saturation effect of the preamplifier becomes severe as the pulse energy increases, Eq. (16) should be used to calculate the average input pulse power $P_{ave}$ for Eq. (12).

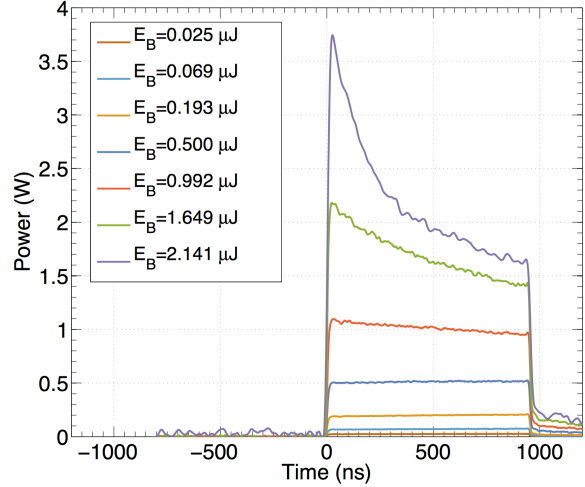

Fig. 10 Input pulse shapes for the 1500 m 9/125 fiber at different pulse energies

The input and output spectra at different pulse energies are shown in Fig. 11. As the blue curves show, with increased pulse energy, the input bandwidth decreases due to the spectral gain narrowing effect. After propagation through the 1500 m 9/125 passive fiber, the spectral bandwidth is broadened due to self-phase modulation, as can be seen from the yellow curves in Fig. 11.

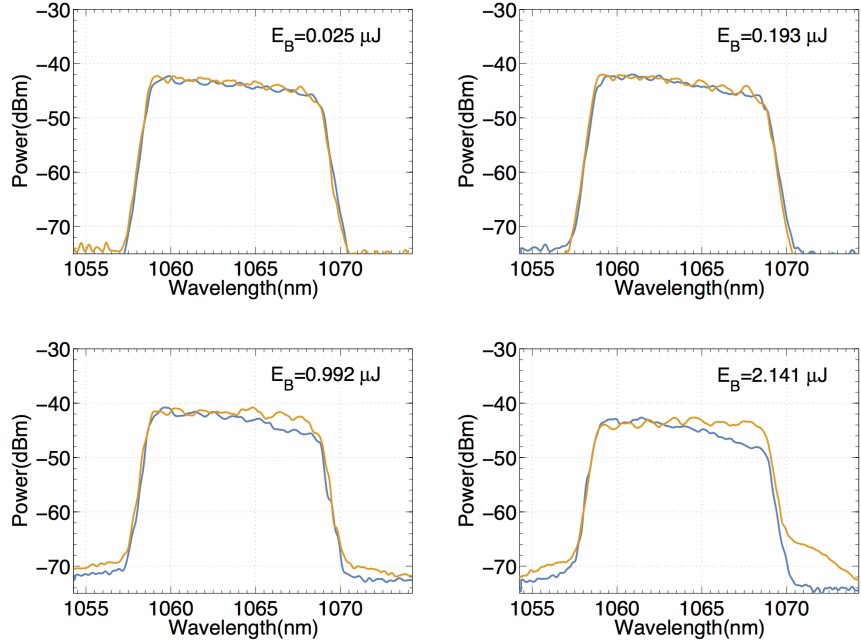

Fig. 11 Input (blue curve) and output (yellow curve) spectra for the 1500 m 9/125 fiber at different pulse energies.

In this experiment, $L_{NL} > 140$ m and $L_D < 6$ m, thus the optical wave is in the weakly nonlinear regime ($L_D \ll L_{NL}$). By employing Eq. (12)(rectangular) and the related equations, the calculated spectral broadening of the pulses at different input pulse energies are obtained. The input and output bandwidth versus calculation results by our model are shown Fig. 12. The purple dashed line by our model gives qualitative predictions for the experimental results with maximum error of 6% and mean error of 3%.

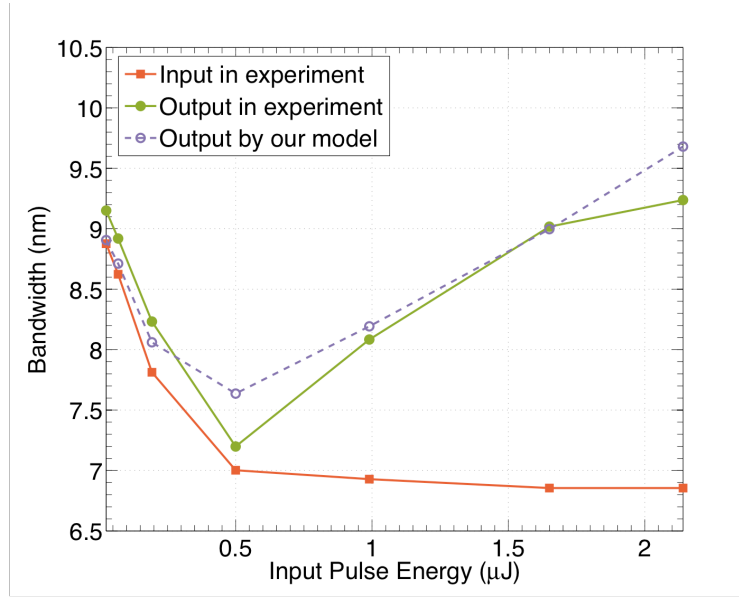

Fig. 12 Experimental results versus calculation results by our model for the broadband incoherent light in the 1500 m 9/125 passive fiber

### 3.2 Experiments for fiber amplifiers

Spectral broadening of the narrowband incoherent light in the gain fibers is also investigated. The experimental setup is shown in Fig. 13. The pulse duration of the incoherent seed light is set to 80 ns. The output pulse from the preamplifiers is further amplified by another two stages of SM fiber amplifiers, each with a fiber grating placed before to narrow down the spectrum. Then the pulse is sent into an 8-nm band-pass filter to filter out the ASE, and then amplified in a 4 m 15/130 Yb-doped double clad fiber amplifier. The gain fiber has a core diameter of 15.6 µm, numerical aperture of 0.084 and the effective modal area of 213 µm$^2$.

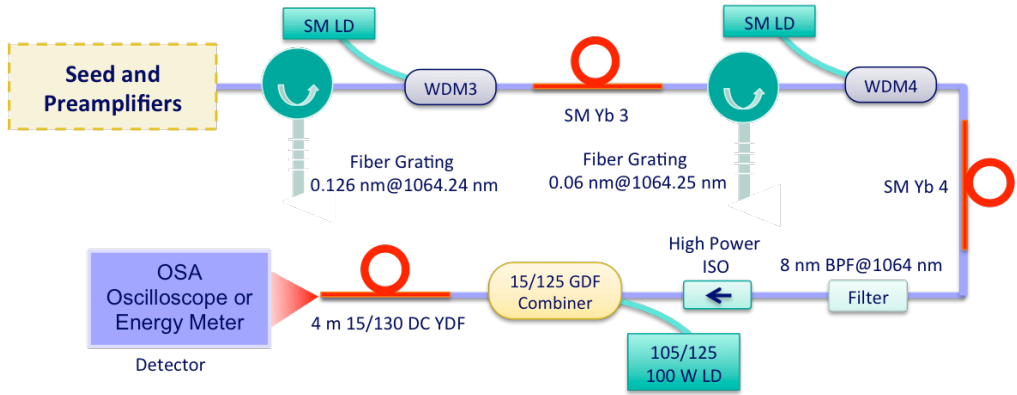

Fig. 13 Experimental setup for spectral broadening of narrowband incoherent light in the fiber amplifier

The seed for the 15/130 Yb-doped fiber amplifier has a 3-dB bandwidth of 0.114 nm, pulse energy of 3.2 µJ and peak power of 40 W. The pulse energy can be amplified to 250 µJ before energy saturates. The output pulse shapes at different output energies are shown in Fig.

14. Compared with the input pulse with a close to flat-top shape as shown in the inset, the output pulse is distorted severely due to the strong gain saturation effect.

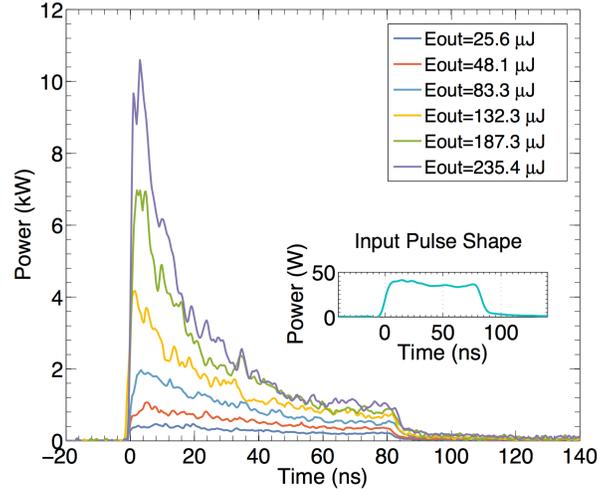

Fig. 14 Pulse shapes from the 15/130 fiber amplifier at different output energies. Inset: input pulse shape

The spectrum of the seed for the 15/130 fiber amplifier is shown in Fig. 15 (a). The seed has a clean spectrum with 3-dB bandwidth of 0.114 nm. The output spectra at different pulse energies are shown in Fig. 15 (b). When the output energy is less than 132 µJ, the light pulse has a broadened but clean spectrum. When pump power is further increased and the output energy is more than 187 µJ, the output spectrum begins to show the four-wave-mixing induced spectrum broadening located mostly at the longer wavelength region.

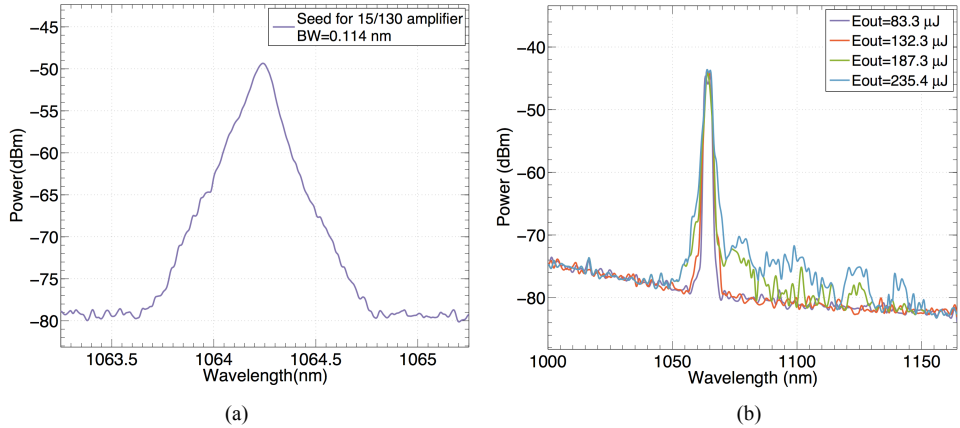

(a) (b)

Fig. 15 (a) Spectrum of the seed for the 15/130 fiber amplifier. (b) Spectra from the 15/130 fiber amplifier

In this experiment, $L_{NL} < 27$ m and $L_D \approx 8500$ m, thus the optical wave is in the strong nonlinear regime ($L_{NL} \ll L_D$). By applying Eq. (12)(Gaussian) and the related equations, the spectral broadenings of the pulses at different output energies are calculated. Since there is strong gain saturation effect in this fiber amplifier, calculations with gain saturation effect are adopted. The output pulse shape is employed to calculate $P_{ave}$ in Eq. (16). Fig. 16 shows the experimental input and output bandwidth and the calculation results by our model. Obviously, our model gives the bandwidth predictions qualitatively for the experiments in fiber amplifiers, with mean error of less than 10 %.

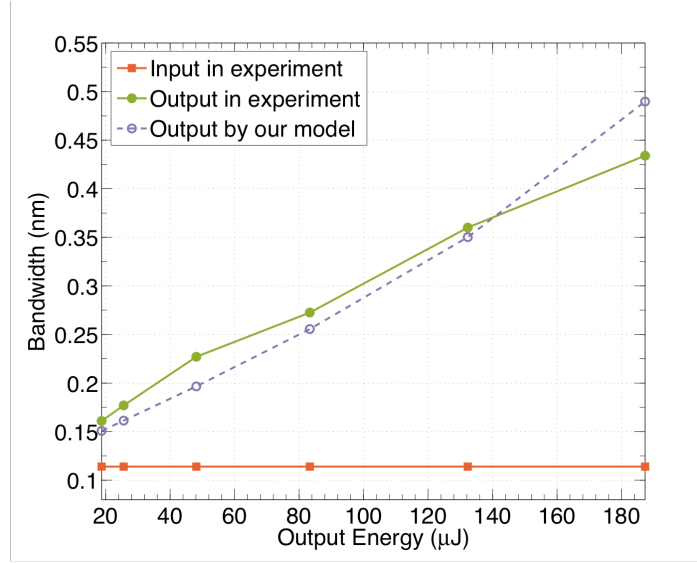

Fig. 16 Experimental results vs. calculation results by our model for the narrowband incoherent light in the 15/130 DC YDF fiber amplifier

*3.3 Discussions*

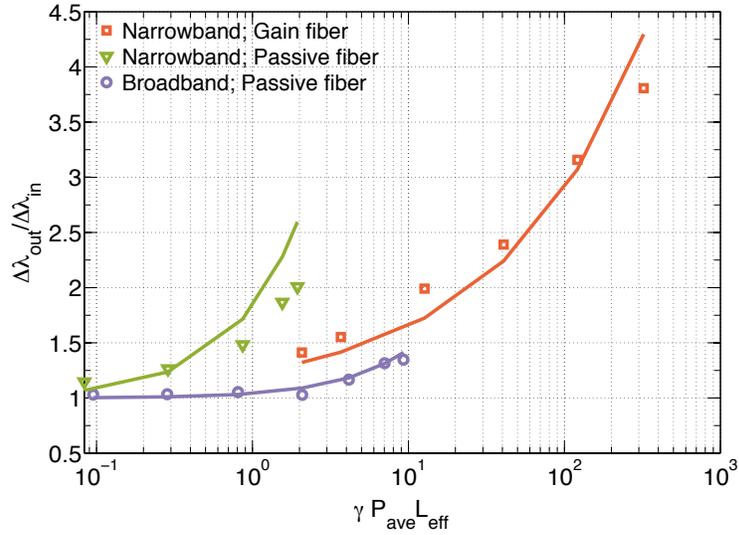

Fig. 17 Experimental results (scattered points) and predictions by our model (solid lines) for the above three series of experiments

For the above three series of experiments, the output-input bandwidth ratio $\Delta\lambda_{out}/\Delta\lambda_{in}$ versus the parameter $\gamma P_{ave} L_{eff}$ is shown in Fig. 17. The scattered points and solid lines are the experimental data and the predictions by our model (twice-calculation method), respectively. The red points and lines are for the narrowband spectral broadening in the 4 m 15/130 fiber amplifier ($L_{NL} \ll L_D$), where our model provides predictions with mean error of less than 10% compared with the experimental results. The green ones are for the narrowband spectral broadening in the 500 m 9/125 passive fiber ($L_{NL} \ll L_D$), where our model agrees qualitatively with the experimental results, with mean errors of less than 16%. The purple ones are for the broadband spectral broadening in the 1500 m 9/125 passive fiber ($L_D \ll L_{NL}$), where our model matches well with the experimental results, with mean error of less than 3%. In conclusion, our model achieves qualitative agreements with the experimental results, no matter for narrowband or broadband incoherent light, in passive or active fibers (as long as spectral gain narrowing effect is negligible in the active fibers), regardless of the dominance of Kerr nonlinear effects or dispersive effects.

One drawback of this model is that it cannot provide the spectral shape evolution of the incoherent light. Nonetheless, it gives a qualitative and readily description of incoherent light spectral broadening, which is proper and convenient in applications where the rough estimations of spectral broadening ratios are required.

## 4. Conclusion

In this paper, an intuitive phenomenological model for the spectral broadening of incoherent light in both passive fibers and fiber amplifiers is presented. The incoherent light is approximated by a cosine power-modulated light with the modulation period depending on the coherent time and the modulation depth depending on the dispersion factor in fibers. A simple spectral broadening formula is given and the twice-calculation method is put forward, which are applicable in the strong nonlinear regime as well as in the weakly nonlinear regime. To verify this method, both the straightforward NLSE-based numerical modeling and the verification experiments are performed. Our model achieves qualitative agreements with the NLSE-based numerical modeling and the experimental results. This model provides us a simple and practical tool to calculate the spectral broadening ratio of incoherent light in fibers.

**Acknowledgments**

This research was supported by the National High Technology Research and Development Program of China, the National Natural Science Foundation of China (Grant No. 61475081), and the State Key Laboratory of Tribology, Tsinghua University (SKLT2014B09).